\documentclass[10pt,aps,twocolumn,showpacs,amsmath,amssymb,eqnarray,array]{revtex4}
\usepackage{anysize}
\marginsize{1.9cm}{1.9cm}{1.55cm}{1.55cm}
\usepackage{graphicx}
\usepackage{dcolumn}
\usepackage{bm}
\begin{document}


\title{Quantum Pinch Effect\\}
\hspace{2.0cm}
\author{Manvir S. Kushwaha}
\affiliation
{\centerline {Department of Physics and Astronomy, Rice University, P.O. Box 1892, Houston, TX 77251, USA}}
\date{\today}

\begin{abstract}

We investigate a two-component, cylindrical, quasi-one-dimensional quantum plasma subjected to a
{\em radial} confining harmonic potential and an applied magnetic field in the symmetric gauge.
It is demonstrated that such a system as can be realized in semiconducting quantum wires offers
an excellent medium for observing the quantum pinch effect at low temperatures. An exact
analytical solution of the problem allows us to make significant observations: surprisingly, in
contrast to the classical pinch effect, the particle density as well as the current density
display a {\em determinable} maximum before attaining a minimum at the surface of the quantum wire.
The effect will persist as long as the equilibrium pair density is sustained. Therefore, the
technological promise that emerges is the route to the precise electronic devices that will control
the particle beams at the nanoscale.

\end{abstract}

\maketitle

\newpage
The pinch effect is one of the most fascinating phenomena in plasma physics with immense applications
to the problems of peace and war. It is the manifestation of radial constriction of a compressible
conducting plasma [or a beam of charged particles] due to the magnetic field generated by the parallel
electric currents. Cylindrical symmetry is central to the realization of the effect. In the literature,
the phenomenon is also referred to as self-focusing, magnetic pinch, plasma pinch, or Bennett pinch.
The pinch effect in the cylindrical geometry is classified after the direction in which the current
flows: In a $\theta$-pinch, the current is azimuthal and the magnetic field axial; in a $z$-pinch, the
current is axial and the magnetic field azimuthal; in a screw (or $\theta$-$z$) pinch, an effort is made
to combine both $\theta$-pinch and $z$-pinch [see Fig. 1].

While a popular reference dates back to 1790 when Martinus van Marum in Holland created an explosion by
discharging 100 Leyden jars into a wire, the first formal study of the effect was not undertaken until
1905 when Pollock and Barraclough [1] investigated a compressed and distorted copper tube after it had
been struck by lightning. They argued that the magnetic forces due to the large current flow could have
caused the compression and distortion. Shortly thereafter, Northrupp published a similar but
independent diagnosis of the phenomenon in liquid metals [2]. However, the major breakthrough in the
topic came with the publication of the derivation of the radial pressure balance in a static z-pinch by
Bennett [3]. Curiously enough, the term ``pinch effect" was only coined in 1937 by Tonks in his work on
the high current densities in low pressure arcs [4]. This was followed by a \emph{first} patent for a
fusion reactor based on a toroidal z-pinch submitted by Thompson and Blackman [5]. The subsequent
progress -- theoretical and experimental -- on the pinch effect in the gas discharges was driven by the
quest for the controlled nuclear fusion.

\begin{figure}[htbp]
\includegraphics*[width=6.5cm,height=7.5cm]{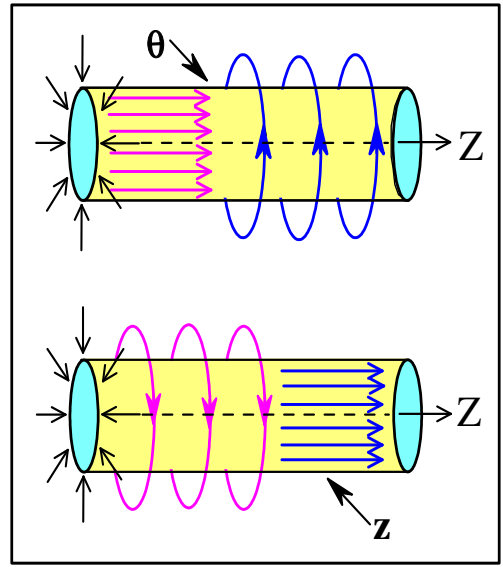}
\caption{(Color online) The schematics of the $\theta$-pinch (the upper panel) and the $z$-pinch
(the lower panel). The curves in blue (magenta) refer to the current (magnetic field).}
\label{fig1}
\end{figure}

Needless to say, the plasma commands a glamorous status in the physics literature by representing the
{\em fourth} state of matter. Our key interest here is in the solid state plasmas (SSP), which share
several characteristic features with but are also known to have quantitative differences from the
gaseous space plasmas (GSP). The GSP is ill-famed for two inherently juxtaposing characteristics:
confinement and instability. The SSP is, on the contrary, a uniform, equilibrium system tightly bound
to the lattice where the boundary conditions make no sense, whereas in GSP they can be crucial. Notice
that only two-component, semiconductor SSP with approximately equal number of electrons and holes can
support a significant pinching. For a single mobile carrier the space-charge electric field due to any
alteration in the charge-density will impede such inward motion. Thus a metal with only electrons
contributing to the conduction is a poor option. The 1960's had seen considerable experimental [6-15]
and theoretical [16-19] efforts focused on studying pinch effect in the bulk SSP, with instabilities
manifesting themselves as voltage and current oscillations at frequencies up to 50 MHz.


Notwithstanding the SSP had been imbued to a greater extent by the research pursuit for the pinch effect
in the GSP, the subject did not gain sufficient momentum because the early 1970's had begun to offer the
condensed matter physicists with the new venues to explore: the semiconducting quantum structures with
reduced dimensions such as quantum wells, quantum wires, quantum dots, and their periodic counterparts.
The continued tremendous research interest in these quasi-n-dimensional electron gas [Q-nDEG] systems can
now safely be accredited to the discovery of the quantum Hall effects -- both integral [20] and fractional
[21]. The fundamental issue behind the excitement in these man-made nanostructures lies in the fact that
the charge carriers exposed to external probes such as electric and/or magnetic fields can exhibit
unprecedented quantal effects that strongly modify their behavior characteristics [22]. To what extent
these effects can influence the behavior of a quantum wire [or, more realistically, Q-1DEG system] in the
cylindrical symmetry, which offers a quantum analogue of the classical structure subjected to investigate
the pinch effect in the conventional SSP [or GSP], has not, to our knowledge, been explored. This is the
motivation behind the present work.

We consider a two-component, quasi-1DEG system characterized by a radial harmonic confining potential and
subjected to an applied (azimuthal and axial) magnetic field in the cylindrical symmetry. The two-component
Q-1DEG systems are comprised of both types of charge carriers [i.e., electrons and holes] and are known to
have been fabricated in a wide variety of semiconducting quantum wires by optical pumping techniques [23-25].
In the linear response theory, the resultant system is characterized by the single-particle Hamiltonian
$H=H_0+H_1$, where
[after Peierls substitution ${\boldsymbol p}\rightarrow ({\boldsymbol p}\pm\frac{e}{c}{\boldsymbol A})$,
with ${\boldsymbol A}={\boldsymbol A}_0 +{\boldsymbol A}_1$]
\begin{align}
H_0 =& \frac{1}{2}m_{e}{\boldsymbol v}^2_{e} + \frac{1}{2}m_{h}{\boldsymbol v}^2_{h} +
       \frac{1}{2}m_{e}\omega^2_{e} r^2+\frac{1}{2}m_{h}\omega^2_{h}r^2\, , \\
H_1 =& \frac{e}{2 c}[({\boldsymbol v}_{e}\!\cdot\!{\boldsymbol A_1} +
                      {\boldsymbol A_1}\!\cdot\!{\boldsymbol v}_{e})
              \! -\! ({\boldsymbol v}_{h}\!\cdot\!{\boldsymbol A_1} +
                      {\boldsymbol A_1}\!\cdot\!{\boldsymbol v}_{h})]\, ,
\end{align}
to first order in ${\boldsymbol A}_1$. Here $c$ is the speed of light in vacuum and $-e (+e)$, $m_e (m_h)$, and
$\omega_e (\omega_h)$ are, respectively, the electron (hole) charge, mass, and the characteristic
frequency of the harmonic oscillator.
${\boldsymbol v}_i=\frac{1}{m_i}\,({\boldsymbol p}\pm \frac{e}{c}\,{\boldsymbol A}_0)$ is the velocity operator
for an electron (hole). A few significant aspects of
the formalism are:
(i) the formal Peierls substitution, by which a magnetic field is introduced into the Hamiltonian, is a direct
consequence of gauge invariance under the transformation $\psi \rightarrow \psi e^{i\phi}$, with $\phi$ being
an arbitrary phase,
(ii) in the Coulomb gauge $\nabla \cdot {\boldsymbol A}=0 \Rightarrow {\boldsymbol A}\cdot {\boldsymbol p}=
{\boldsymbol p}\cdot {\boldsymbol A}$, and
(iii) we choose symmetric gauge, which is quite popular in the many-body theory, defined by
${\boldsymbol A}_0 = B (0, \frac{1}{2} r, 0)$.

\begin{figure}[htbp]
\includegraphics*[width=8cm,height=9cm]{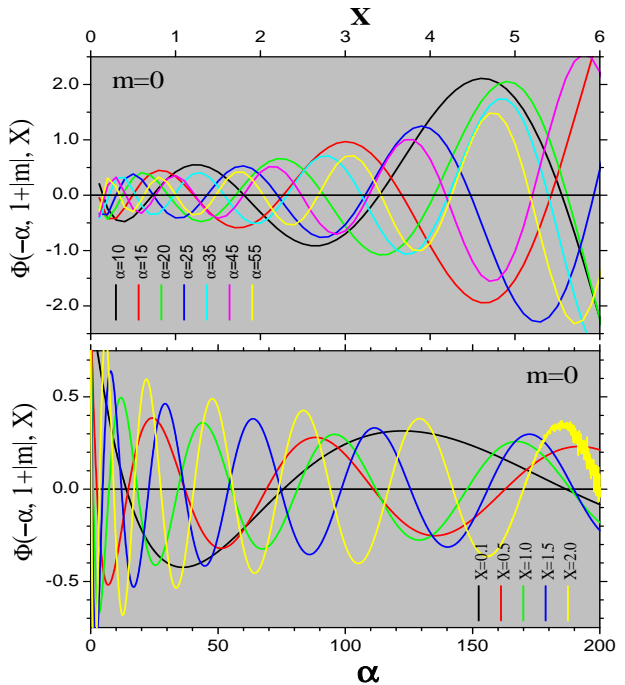}
\caption{(Color online) The plot of the CHF $\Phi (\alpha, 1+|m|, X)$  vs. $X$ (the upper panel) and
$\alpha$ (the lower panel), for $m=0$.}
\label{fig2}
\end{figure}

Next, we consider, for the sake of simplicity, the quantization energy for the electrons equal to that for the
holes implying that $\omega_e=\omega_o=\omega_h$. Then, in the cylindrical coordinates [($r, \theta, z$)], the Schr\"odinger equation $H_0\psi=\epsilon \psi$ for a quantum wire of radius $R$ and length $L$ is solved to
characterize the system with the eigenfunction $\psi (r,\theta,z)=\psi(r)\,\psi(\theta)\,\psi(z)$, where
\begin{align}
\psi(z)&=\sqrt{\frac{2}{L}}\,\sin(k\,z), \,\,\,\,\,\,\,{\rm with \,\,\,k=n\pi/L}\, ,\\
\psi(\theta)&=\sqrt{\frac{1}{2\pi}}\,\, e^{i\,m\,\theta}\, ,\\
\psi(r)&=\frac{N}{s^{|m|/2}}\,e^{-X/2}\, X^{|m|/2}\,\Phi(-\alpha; 1+|m|; X) ,
\end{align}
where $n$ is an integer, $m=0,\pm 1, \pm 2, ...$ is the azimuthal quantum number, $s=\sqrt{m_+/\mu}$, and $X= r^2/(2\ell^2_c)$, with $\ell_c=\sqrt{\hbar/\mu \Omega_+}$ as the effective magnetic length in the problem,
$m_+=m_e+m_h$ as the total mass, $\mu=m_e m_h/(m_e + m_h)$ as the reduced mass, $\Omega_+=s \Omega=s \sqrt {\omega_{ce}\,\omega_{ch}+4 \omega_o^2}$ as the effective hybrid frequency, and $\omega_{ci}=eB/(m_i c)$ as
the cylotron frequency. In Eq. (5), $N$ is the normalization coefficient to be determined by the condition
$\int^{R}_{0} dr\, r \mid \psi (...) \mid^2=1$, which yields
\begin{equation}
N^{-2}=\frac{\ell^2_c}{s^{\mid m \mid}}\!\int^{\zeta}_{0}\!\! dX\,e^{-X} X^{\mid m \mid}
                                              \big [\Phi(-\alpha; 1+|m|; X)\big]^2
\end{equation}
where the upper limit $\zeta=R^2/2\ell^2_c$. In Eq. (5), $\Phi(-\alpha; 1+|m|; X)$ is the confluent
hypergeometric function (CHF) [26], which is a solution of the Kummer's equation:
$X U^{''}\!+[1+\!\!|m|\!\!-X]\,U'\!+\alpha\,U=0$, where
$\alpha=\frac{1}{\hbar \Omega_+}[\epsilon -\frac{\hbar^2 k^2}{2\mu}-\frac{m}{2}\hbar \Omega_-]-
\frac{1}{2}(1+|m|)$, with
$\Omega_-=(\omega_{ce}-\omega_{ch})$ being the effective cyclotron frequency. The wavefunction $\psi (...)$
obeys the boundary condition $\psi (r=R)=0 \Rightarrow \Phi(-\alpha; 1+|m|; \zeta)=0$, which yields
\begin{equation}
\frac{\epsilon'}{\epsilon_r}=\zeta\,
                  \big [\alpha +\frac{1}{2}\,\big (1+m\,\frac{\Omega_-}{\Omega_+}+\mid m \mid \big)\big]
\end{equation}
where $\epsilon'=\epsilon-\frac{\hbar^2 k^2}{2\mu}$ and $\epsilon_r=2\hbar^2/(\mu R^2)$. It is needless to
state that the term $\frac{\hbar^2 k^2}{2\mu}$ is only additive and hence inconsequential. What is
important to notice though is that, for a given $m$, $\zeta$ and $\alpha$ are determined self-consistently
where $\Phi(-\alpha; 1+|m|; X)=0$. Figure 2 demonstrates that $\Phi (...)$ is unambiguously a well-behaved
function over a wide range of $X$ and $\alpha$ and that its period is seen to be decreasing with increasing
$X$ or $\alpha$ as the case may be. Since the rest of the illustrative examples require the quantum wire to
be specified, we choose GaAs/Ga$_{1-x}$Al$_x$As system: $m_e=0.067 m_0$, $m_h=0.47 m_0$, and the aspect
ratio $r_a=L/R=10^3$ [27]. Another important parameter in the problem is the reduced magnetic flux
$\phi/\phi_0$: the ratio of the magnetic flux [$\phi=B\pi R^2$] to the
quantum flux [$\phi_0=h c/e$] -- yielding $\phi/\phi_0=\zeta$.

\begin{figure}[htbp]
\includegraphics*[width=8cm,height=9cm]{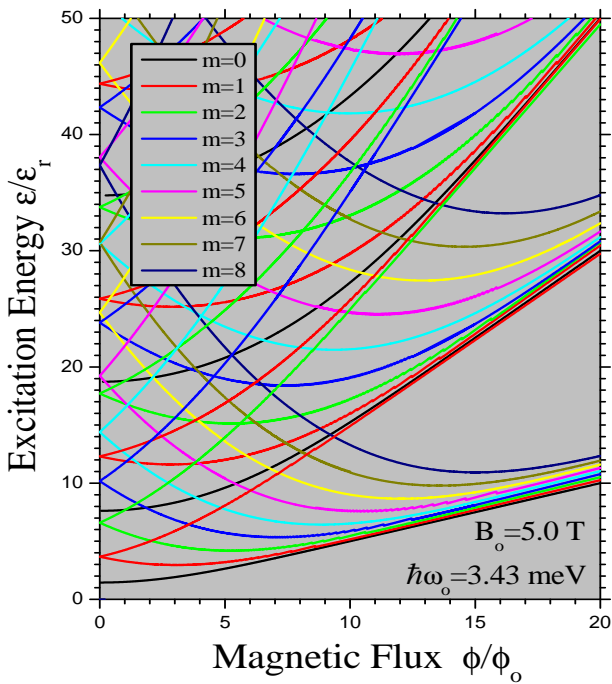}
\caption{(Color online) The excitation spectrum for a (finite) cylindrical quantum wire subjected to
a magnetic field in the symmetric gauge. The y (x) axis refers to the reduced energy (reduced
magnetic flux). The lower (upper) arm of a wedge corresponds to a negative (positive) value of $m$.}
\label{fig3}
\end{figure}

Figure 3 illustrates the excitation spectrum for the cylindrical quantum wire of finite length as a function
of the reduced magnetic flux $\phi/\phi_0$, for several values of $m$. This is a result of (involved)
computation based on Eq. (7) within the strategy as stated above. The parameters used are as listed inside
the picture. For every $m$, the lower (upper) branch of the wedge corresponds to the negative (positive) value
of $m$. It is interesting to see that both arms of each wedge gradually approach the Landau levels at higher
magnetic flux, just as in the case of the Fock-Darwin spectrum of a quantum dot [see, e.g., Ref. 22]. Notice
that this is not a mere coincidence, rather a consequence of the isomorphism in the expression of the
single-particle energies for the two quantum systems.

Next, we proceed to solve the time-dependent Schrodinger equation in order to derive the equation of continuity:
$\frac{\partial \rho}{\partial t}+\nabla \cdot {\boldsymbol J}=0$, where the electrical current density
${\boldsymbol J}$ is given by [with $m_e < m_h$]
\begin{equation}
{\boldsymbol J}=\frac{i e \hbar}{2 m_e}\,\big (\psi \nabla \psi^{*}-\psi^{*}\nabla \psi \big ) +
\frac{e^2}{m_e c}\,{\boldsymbol A}\,\psi^{*}\psi\, ,
\end{equation}
where the vector potential ${\boldsymbol A}\,\, [={\boldsymbol A}_0+{\boldsymbol A}_1]$ is approximated such that
we seek to measure its {\em ac} part (${\boldsymbol A}_1$) just as we do its {\em dc} one (${\boldsymbol A}_0$),
i.e., we express ${\boldsymbol A}_0$ in the symmetric gauge just as before and ${\boldsymbol A}_1 = B (0, 0, r)$.
This is not to say that there aren't other, more subtile, ways to complicate the situation, but since this is
the first paper of its kind, we choose to stick to the bare-bone simplicity -- the complexity will (and should)
come later. Consequently, it is not difficult to split Eq. (8) into its scalar components. The result is
\begin{align}
J_{\theta}&=\Big[\frac{e \hbar m}{m_e r} + \frac{e^2 B r}{2 s m_e} \Big]\,
               \big| \psi (r, \theta, z)\big|^2\nonumber\\
J_{z}&= \frac{e^2 B r}{s m_e}\, \big| \psi (r, \theta, z)\big|^2\, .
\end{align}
The very nature of the magnetic field -- with its charismatic role in localizing the charge carriers in the
plane perpendicular to its orientation -- elucidates why the radial component of the current density $J_{r}=0$.
It is interesting to notice that in the absence of the $z$ component of ${\boldsymbol A}_1$, $J_z$ will be zero
for a quantum wire of {\em finite} length (as is the case here). This is contrary to the case of a wire of an
{\em infinite} length -- where $J_z \ne 0$ even when $[{\boldsymbol A}_1]_z=0$. Equations (9) should, in
principle, provide the clue to the quest in the problem.

\begin{figure}[htbp]
\includegraphics*[width=8cm,height=9cm]{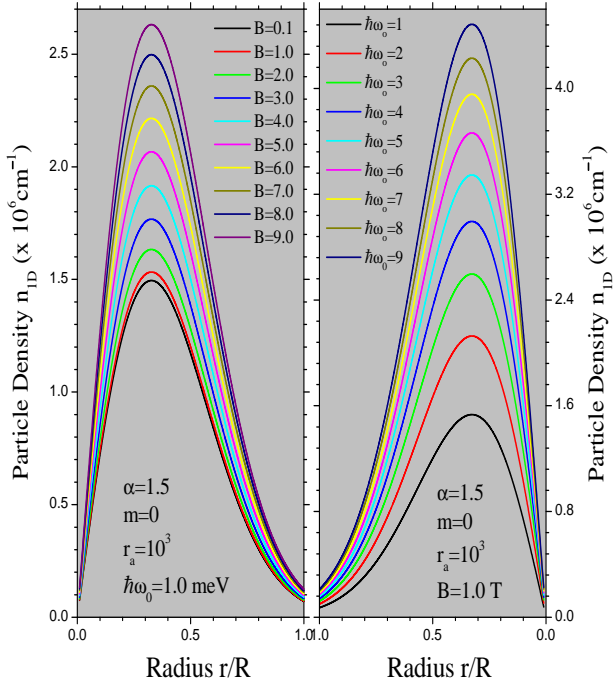}
\caption{(Color online) The particle density $n_{1D}$ as a function of the reduced radius $r/R$ for the
several values of the magnetic field (left panel) and confinement potential (right panel). The parameters
are as listed inside the picture. Again, it is a GaAs/Ga$_{1-x}$Al$_{x}$As quantum wire.}
\label{fig4}
\end{figure}

Figure 4 shows the 1D particle density ($n_{1D}$) in the quantum wire as a function of the reduced radius for
various values of the magnetic field (left panel) and confinement potential (right panel). The confinement
potential in the left panel is $\hbar \omega_0=1.0$ meV and the magnetic field in the right panel is $B=1.0$ T.
The Other parameters are listed inside the picture. The particle density shows a maximum. We were able to find
out that this maximum occurs unequivocally at
$\frac{d}{d r}[n_{1D}]=\frac{d}{d r} [r \Phi^2 (\alpha, 1+|m|, X)]=0\Rightarrow r/R=1/\sqrt{2\,\zeta}=0.3271$.
Since the particle density is fundamental to most of the electronic, optical, and transport properties, we
expect the current features to make a mark in those phenomena, at least, in the cylindrical symmetry. In a
quantum system the particle density has much to with the Fermi energy in the system. For the fact that each
quantum level can take two electrons with opposite spin, the Fermi energy $\epsilon_F$ of a system of
$\mathcal{N}$ noninteracting electrons for a {\em finite} system at absolute zero temperature is equal to the
energy of the $\frac{1}{2}\mathcal{N}$-th level. As such, one can compute self-consistently the Fermi energy
in a {\em finite} quantum wire containing $\mathcal{N}$ electrons through this expression:
$\mathcal{N}=2 \sum _{m}\theta (\epsilon_F - \epsilon'_{m})$, where $\theta(...)$ is the Heaviside step
function and the summation is only over the azimuthal quantum number $m$. The prime on $\epsilon_{m}$ has the
same meaning as explained before. Because the electronic excitation spectrum happens to be so very intricate
[see Fig. 3], the Fermi energy will not be a smooth function of the magnetic field (or flux). This intricacy
should also lead to complex structures in the magnetotransport phenomena.

Figure 5 depicts the current density as a function of the reduced radius for several values of the magnetic field.
The parameters used are: the confinement potential $\hbar \omega_0=1.0$ meV, aspect ratio $r_a=1000$,
$\alpha=1.5$, and $kz=1.8^o$. Other material parameters are the same as before. We observe that the larger the
magnetic field, the greater the current density. This is what we should expect intuitively: the larger the
magnetic field, the stronger the confinement of the charge carriers near the axis and hence greater the current
density. The most important aspect this figure reveals is that there is a maximum in the current density and this
maximum is again defined exactly by $r/R=0.3271$. In a certain way, Fig. 4 substantiates the features observed in
Fig. 5. This tells us that in a quantum wire the maximum of charge density lies at $r/R=0.3271$ instead of exactly
at the axis. Note that the classical pinch effect in conventional (3D) SSP [6-19] does not share any such feature.
Very close to the axis and to the surface of the quantum wire, the minimum of the current density is smallest but
still nonzero. Traditionally, $\theta$-pinches ($\Rightarrow J_{\theta}$) tend to resist the plasma instabilities
due to the famous {\em frozen-in-flux} theorem [28], whereas $z$-pinches ($\Rightarrow J_z$) tend to favor the
confinement phenomena. The magnitude of $J_{z}$ is double of that of the $J_{\theta}$ just as dictated by Eqs. (9).
We believe that these currents are of moderate strength and would not cause an undue heating of the two-component
plasma in the quantum wires which are generally subjected to experimental observations at low temperatures.

There are several other important (and interrelated) issues such as the equilibrium, temperature, recombination,
and population inversion, which would certainly give a better insight into the problem. The aspect of population
inversion enabling the magnetized quantum wires to act as optical amplifiers has recently been discussed in a
different context [29]. These issues are deferred to a future publication.

\begin{figure}[htbp]
\includegraphics*[width=8cm,height=9cm]{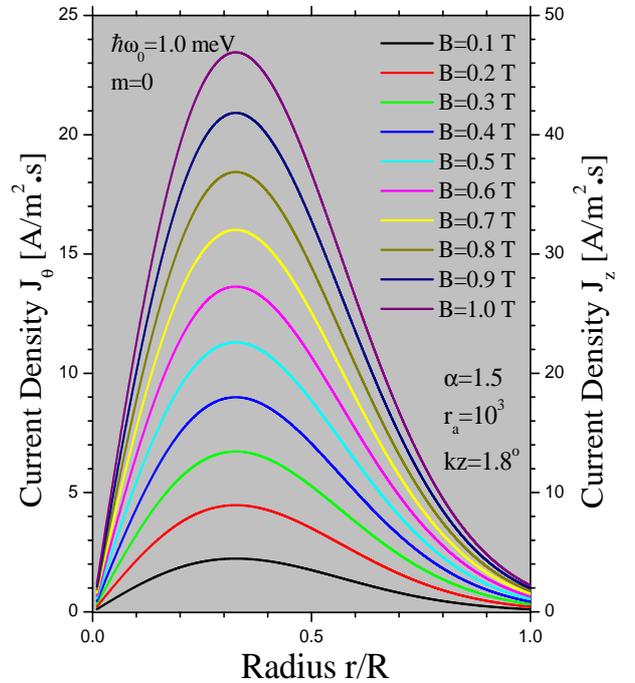}
\caption{(Color online) The current density $J_{\theta}$ ($J_{z}$) on the left (right) vertical axis
vs. the reduced radius $r/R$ for the several values of the magnetic field. The parameters used are:
$\hbar \omega_0=1.0$ meV, aspect ratio $r_a=1000$, $\alpha=1.5$, and $kz=1.8^o$. We consider a
GaAs/Ga$_{1-x}$Al$_{x}$As quantum wire just as before.}
\label{fig5}
\end{figure}


In summary, we have investigated the quantum analogue of the classical pinch effect in finite quantum wires with
cylindrical symmetry. Since the late 1940s the pinch effect in a gas discharge has been investigated intensively
in laboratories throughout the world, because it offers the possibility of achieving the magnetic confinement of
a hot plasma (a highly ionized gas) necessary for the successful operation of a thermonuclear or fusion reactor.
In solid state plasmas the issues related to confinement, as discussed above, are not encountered. However, since
no system provides an ideal environment in the real world, the SSPs also pose challenges. In a conventional SSP
with no impurities, the thermally excited pair density is a function of temperature only. Any deviation from this (thermal) equilibrium can (and, generally, does) give rise to recombination, which, in turn, affects the pinching
process. The desired maintenance of the equilibrium pair density occurs through various processes such as Coulomb interactions and particle-lattice interactions, which are not independent and hence cause unintended consequences.
The quantum wires are the systems in which most of the experiments are performed at low (close to zero)
temperatures. Therefore, the risks of thermal non-equilibrium and recombination are much smaller than those
ordinarily encountered in conventional SSP. This implies that two-component quantum wires at low temperatures
offer an ideal platform for the realization of the quantum pinch effect.

Myriad of applications of quantum pinches bud out from the very thought of the self-focused, two-component plasma
in a quantum wire with cylindrical symmetry. The plasma by definition is electrically conductive implying that it
responds strongly to electromagnetic fields. The self-focusing (or pinching) only adds to the response. The
quantum wire brings all this to the nanoscale. Therefore, we are designing an electronic device that can (and
will) control the particle beams at the nanoscale. Potential applications include extremely refined nanoswitches,  nanoantennas, optical amplifiers, and precise particle-beam nanoweapons, just to name a few. The greatest
advantage of the quantum pinch effect over its classical counterpart is that it offers a Gaussian-like cycle of
operation with two minima passing through a maximum. The smooth functionality of the plasma devices is, however,
based on a single tenet: there must be means not only to produce it, but also to sustain it.

The author feels grateful to L.N. Pfeiffer who affirmed the feasibility of achieving the aspect ratio
of $r_a=1000$ in the semiconducting quantum wires within the current technology and for the fruitful
discussions. I would also like to thank H. Sakaki, Naomi Halas, and Peter Nordlander for stimulating
discussions on various aspects related to the subject. He is appreciative of Kevin Singh for the
invaluable help with the software during this investigation.


\end{document}